\documentclass[11pt,a4paper]{article} 
 
        \usepackage[utf8]{inputenc}
        \usepackage{fontenc}
	\usepackage[italian,english]{babel}
	\usepackage{amsmath}
	\usepackage{geometry}
	\usepackage{amsthm}
	\usepackage{mathrsfs}
	\usepackage{bbold}
	\usepackage{mathtools}
	\usepackage{amsfonts}
	\usepackage{graphicx}
	\usepackage{setspace}
	\usepackage{physics}
	 \usepackage{comment}
	 \usepackage{bbm}
	  \usepackage{layout}
	\usepackage{xcolor}
     \usepackage[big]{layaureo}
	 \usepackage{dsfont}	
	 \usepackage{slashed}
 	 \usepackage{amssymb}
	 \usepackage{tensor}
	\usepackage{fancyhdr} 
 	\usepackage{tikz-cd}
 	\usepackage{braket}
 	\usepackage{tikz}
 	\usepackage[compat=1.1.0]{tikz-feynman}
 	\usepackage{subfigure}
	\usepackage{fancyhdr} 
	\usepackage{booktabs}
	\usepackage{braket}
	
\usepackage{jheppub} 

\usepackage{natbib}
\title{Generating functions for Higgs/Coulomb branch operators from 1d-3d cohomological equivalence}

\author[a]{Luigi Guerrini,} 
\author[b]{Silvia Penati,}
\author[c]{Itamar Yaakov}

\affiliation[a]{Dipartimento SMFI, Universit\`a di Parma and INFN Gruppo Collegato di Parma, Viale G.P. Usberti 7/A, 43100 Parma, Italy}
\affiliation[b]{ Dipartimento di Fisica, Universit\`a degli studi di Milano--Bicocca, and INFN, Sezione di Milano--Bicocca, Piazza della Scienza 3, I-20126 Milano, Italy }
\affiliation[c]{INFN, sezione di Milano-Bicocca, I-20126 Milano, Italy}

\emailAdd{luigi.guerrini@unipr.it} 
\emailAdd{silvia.penati@mib.infn.it} 
\emailAdd{itamar.yaakov@mib.infn.it}

\abstract{We provide a proof for the conjectured equality of the generating function of integrated Higgs and Coulomb branch topological operators in 3d $\mathcal{N}\ge4$ gauge theories and the three sphere partition function deformed by mass or FI parameters. The equality is the result of cohomological equivalence and applies to all theories in this class, including ABJM and other generalized Gaiotto-Witten models, and those without an explicit supersymmetric Lagrangian.}

	\newcommand{\beq}{\begin{equation}}
	\newcommand{\bea}{\begin{eqnarray}}
	\newcommand{\eea}{\end{eqnarray}}
	\newcommand{\eeq}{\end{equation}}

	\makeatletter
	\newcommand{\aextp}{\@ifnextchar^\@aextp{\@aextp^{\,}}}
	\def\@aextp^#1{\mathop{\bigwedge\nolimits^{\!#1}}}
	\makeatother
	
	\makeatletter
	\newcommand{\extp}{\@ifnextchar_\@extp{\@extp_{\,}}}
	\def\@extp_#1{\mathop{\aextp\nolimits_{\!#1}}}
	\makeatother

	\theoremstyle{definition}

\begin{document}
\maketitle

\section{Introduction}

Correlation functions of local operators are among the most basic observables in a Quantum Field Theory. Spacetime and global symmetries restrict the functional form of these correlators, and, in some situations, can be sufficiently powerful to determine them up to a finite number of numerical constants. These constants can themselves be constrained, or even calculated exactly, using additional inputs. For example, the conformal bootstrap program uses conformal symmetry, crossing symmetry, and unitarity in order to constrain conformal data, which in turn determines correlators involving an arbitrary number of spacetime points. 

Supersymmetry can likewise be used to constrain, and sometimes calculate, the correlators of local operators which are invariant under a subset of the supersymmetry algebra. In order to exhibit a collection of operators which are all invariant under the same supersymmetries, it may be advantageous to endow the operators with an explicit coordinate dependence \cite{deMedeiros:2001wqm, Drukker:2009sf}. Recently, it was demonstrated that such a collection of operators exists in any 4d $\mathcal{N}\ge2$ superconformal field theory \cite{Beem:2013sza}. The operators in question are restricted to a 2d surface inside the 4d spacetime, and their correlation functions were shown to be governed by the infinite 2d conformal algebra. 

3d theories with $\mathcal{N}\ge 4$ superconformal symmetry were shown to admit an analogous collection of operators which, when arranged along a straight line, are governed by a one dimensional topological field theory \cite{Beem:2013sza, Chester:2014mea, Beem:2016cbd}. The operators in this topological sector come in two types: Higgs branch operators and Coulomb branch operators, depending on whether they are associated to hypermultiplet or vector multiplet excitations. In \cite{Dedushenko:2016jxl, Dedushenko:2017avn, Dedushenko:2018icp}, it was shown that the relevant sectors can be probed even for non-conformal theories which flow to an IR superconformal fixed point, by considering the theory on the three sphere rather than in flat space. The authors showed that the numerical constants associated with the topological sector, i.e. the data of the 1d Topological Quantum Field Theory (TQFT), can be computed, in this case, by using supersymmetric localization. The localization procedure can be extended to other backgrounds. For an explicit example on $S^2\times S^1$ see \cite{Panerai:2020boq}. 

In \cite{Agmon:2017xes, Binder:2019mpb}, the generating function for specific \emph{integrated} correlators in the topological sector was further conjectured to coincide with the three sphere partition function in the presence of certain supersymmetry preserving deformations.  For the Higgs branch operators, the conjecture takes the form
\begin{equation}\label{conjecture}
\hspace{-0.3cm} \Big\langle \int d\varphi_1\, J^{A_1}(\varphi_1)\dots \int d\varphi_n \,J^{A_n}(\varphi_n)\Big\rangle=\left( -\frac{1}{4\pi r^2} \right)^{\! n} 
\, \frac{1}{Z}\frac{\partial^n Z[m_1,\dots ,m_n]}{\partial m^{A_1}\dots \partial m^{A_n}}\bigg|_{m^{A_1} \!, m^{A_2} \!, \cdots = 0}
\end{equation}
where the $J^{A_i}$ are specific gauge invariant Higgs branch operators, and $Z[m_1,...]$ is the three sphere partition function deformed by supersymmetric mass terms, which can also be viewed as the generating function of a different set of local operators integrated over the entire three sphere. The conjecture was proven for a subclass of theories to which it was conjectured to apply, using localization.

In principle, localization is applicable to any Lagrangian field theory with enough supersymmetry. However, certain technical challenges remain when considering theories for which supersymmetry is not realized off-shell \cite{Griguolo:2021rke}. This applies, for example, to 3d $\mathcal{N}\ge 4$ gauge theories with Chern-Simons terms, of the type considered in e.g. \cite{Gaiotto:2008sd, Hosomichi:2008jd}, including the ABJM model and its generalizations \cite{Aharony:2008ug, Aharony:2008gk, Hosomichi:2008jb}. Localization also cannot be directly applied to non-Lagrangian field theories. For these reasons, it would be very useful to prove the conjecture without resorting to localization for the fundamental fields. 

\vskip 10pt
In fact, in this paper we show that the result can be established by exhibiting the following cohomological equivalence
\begin{equation}\label{eq:main}
\pdv{m}\left(S_{\textup{mass}}[m]-4\pi r^2 m\oint_{S^1_{\varphi}}    J(\varphi)   \right)= \acomm{Q_\beta^H}{\dots}\,,
\end{equation}
where $S_{\textup{mass}}[m]$ and $-4\pi r^2 m \oint_{S^1_{\varphi}} J(\varphi)$ are the mass deformations of the 3d and 1d theories respectively, while the rhs is the variation of an appropriate fermionic operator in the theory. Precisely, we prove that the equality between generating functions follows from this cohomological equivalence, and exhibit the appropriate fermionic operator.

In section \ref{sec:topological_operators} we give an overview of the construction of the topological sector for ${\cal N} \geq 4$ supersymmetric gauge theories in three dimensions. In section \ref{sec:cohomological_equivalence_proof} we establish the cohomological equivalence at the level of linear coupling to the background fields. For a particular example involving free hypermultiplets, we prove the validity of \eqref{eq:main} also at non-linear level in $m$. Finally, some conclusions are collected in section \ref{sec:conclusions}, while conventions and useful technical details can be found in three appendices. 

\vskip 5pt
\noindent
Note added: As this paper was nearing completion, \cite{Bomans:2021ldw} appeared, whose last section has significant overlap with our results.

\section{Topological operators: an overview}\label{sec:topological_operators}

In this section we review the construction of topological operators in 3d $\mathcal{N}\ge 4$ supersymmetric field theories \cite{Beem:2013sza, Chester:2014mea, Beem:2016cbd}. These are a particular subset of operators  whose correlation functions, when  localized on a 1d submanifold, are constant. 

We begin by discussing the case of $\mathcal{N}\ge 4$ superconformal field theories (SCFTs). In 3d the $\mathcal{N}=4$ superconformal group is ${\rm Osp}(4|4)$. The corresponding $\mathfrak{osp}(4|4)$ superalgebra contains as its maximal bosonic subalgebra the 3d conformal algebra $\mathfrak{sp}(4)$ and the R-symmetry algebra $\mathfrak{su}(2)_H\oplus\mathfrak{su}(2)_C$, whose generators are denoted by ${R_{a}}^{b}$ and ${{\bar R}_{\dot a}}^{\; \; \dot b}$, respectively. The odd sector contains Poincar\`e supercharges $Q_{\alpha, a \dot a}$ and superconformal charges $S_{\alpha,a\dot a}$.

The essential ingredient for constructing a set of topological operators closed under OPE, is the existence of a nilpotent supercharge $\mathcal{Q}$, which can be obtained as a linear combination of the $Q_{\alpha, a \dot a}$ and $S_{\alpha,a\dot a}$ generators \cite{Beem:2013sza, Chester:2014mea}. This supercharge can be used to define a $\mathcal{Q}$-exact R-twisted translation $\hat P\sim P+R$, where $P$ is the ordinary translation along a 1d submanifold. It then follows that given a $\mathcal{Q}$-cohomology class of operators localized at the origin, they can be translated away from the origin by acting with $\hat P$, without affecting the $\mathcal{Q}$-cohomology. As a consequence, correlation functions of $\mathcal{Q}$-closed operators placed on the 1d submanifold turn out to be piecewise constant, depending at most on the order of the insertions. We call this set of operators the \emph{topological sector}.  

To be concrete, in 3d euclidean flat space we take the 1d submanifold to be the line $x_1=x_2=0$ and focus on the construction of topological operators in the Higgs branch first. The maximal superconformal algebra preserved by the line is $\mathfrak{su}(2|2)$, whose bosonic subalgebra includes the $\mathfrak{so}(2,1)$ conformal algebra and the $\mathfrak{su}(2)$ R-symmetry algebra that we identify with $\mathfrak{su}(2)_H$. The bosonic generators are $P_{3}$, $K_{3}$, $D$, ${R_{a}}^b$, and the fermionic ones are $Q_{1,a\dot 2}$, $Q_{2,a\dot 1}$, $S_{2,a\dot 1}$, and $S_{1,a\dot 2}$. 

If we consider the two supercharges
\begin{equation}\label{eq:supercharges}
Q_1^H\equiv Q_{11\dot{2}}+\frac{1}{2r}{S^2}_{2\dot{2}}\,,\qquad Q_2^H\equiv Q_{21\dot{1}}+\frac{1}{2r}{S^1}_{2\dot{1}}\,,
\end{equation}
with $r$ being an arbitrary length parameter, from the $\mathfrak{su}(2|2)$ algebra it is easy to see that they are both nilpotent operators. Moreover, their anticommutator reads
\begin{equation}
\acomm*{Q_1^H}{Q_2^H}=-M_{12}+R_{\dot 1\dot 2} \equiv Z
\end{equation}
where $M_{12}$ is the generator of rotations in the plane orthogonal to the line. 
It then follows that since $Z$ must vanish on the $Q^H_{1,2}$ cohomology classes, operators belonging to these classes are inserted along the fixed point locus of $M_{12}$ and have zero $R_{\dot 1\dot 2}$ charge. 

We now perform a topological twist by combining the $\{P_3, K_3, D\}$ generators of the $\mathfrak{so}(2,1)$ conformal algebra along the line and the $\mathfrak{su}(2)_H$ R-symmetry generators. It turns out that the twisted generators are all $Q_{1,2}^H$-exact. Therefore, we can first construct operators localized at the origin from $Q_{1,2}^H$-closed, gauge invariant operators of the 3d theory. 
Explicitly, the cohomology of the two supercharges contain local operators $O_{a_1,\,\dots,\,a_n}$ with the following properties \cite{Chester:2014mea}: they are Lorentz scalars, transform in the $(\mathbf{n+1},\,\mathbf{1})$ of  $\mathfrak{su}(2)_H\oplus\mathfrak{su}(2)_R$, and have conformal dimension $\Delta=n/2$. We then move them along the line by applying the twisted translation generator $\hat{P}=P_3+\frac{\mathbbm{i}}{2r}{R_{2}}^{1}$. The corresponding twisted translated operator at position $\vec{x}=(0,\,0,\,s)$ is given by
\begin{equation}
O(s)=e^{i s \hat{P}}O_{1,\dots,1}(0)e^{-i s \hat{P}}=O_{a_1,\dots,a_n}(\vec x)\big|_{\vec x=(0,0,s)} u^{a_1}\dots u^{a_n}\,, \qquad u^a=\left(1,\frac{s}{2r}\right)\,.
\end{equation}
These operators are still $Q_{1,2}^H$-closed and form the Higgs topological sector of the ${\cal N} \geq 4$ SCFT on the line.

\vskip 10pt

To make contact with localization it is convenient to consider Lagrangian theories defined on $S^3$. The crucial point is that $Q_{1,2}^H$ defined in \eqref{eq:supercharges} where now $r$ is the $S^3$ radius, belong to the Poincaré subalgebra $\mathfrak{su}(2|1)_\ell\oplus\mathfrak{su}(2|1)_r$ introduced in \cite{Dedushenko:2016jxl}. The bosonic part of the algebra is $\mathfrak{so}(4)\oplus\mathfrak{u}(1)_\ell\oplus\mathfrak{u}(1)_r$, where $\mathfrak{so}(4)$ is the isometry algebra of $S^3$ and $\mathfrak{u}(1)_\ell\oplus\mathfrak{u}(1)_r$ is the residual R-symmetry, whose generators are linear combinations of $R_H=\frac{1}{2}{h_a}^b{R_b}^a$ and $R_C=\frac{1}{2}\bar{h}\indices{^{\dot{a}}_{\dot{b}}}\bar{R}\indices{^{\dot{b}}_{\dot{a}}}$\footnote{ $h\indices{_a^b}$ and $\bar{h}\indices{^{\dot{b}}_{\dot{a}}}$ are respectively $\mathfrak{su}(2)_H$ and $\mathfrak{su}(2)_C$ matrices, normalized such that $h\indices{_a^c}h\indices{_c^b}=\delta_a^b$ and $\bar{h}\indices{^{\dot{b}}_{\dot{c}}}\bar{h}\indices{^{\dot{c}}_{\dot{a}}}=\delta_{\dot{a}}^{\dot{b}}$.}.
The cohomology algebra on $S^3$ reads
\begin{equation}
\acomm{Q_1^H}{Q_2^H}=\frac{4\mathbbm{i}}{r}\left(P_\tau+R_C+\mathbbm{i}r \zeta\right)\,,
\end{equation}
where $P_\tau$ is the translation along the great circle $S^1_\tau$ on $S^3$ parametrized by $\tau$ (see the metric in eq. \eqref{eq:metric}). We have included the central extension $\zeta$ associated to the presence of a FI term \cite{Dedushenko:2016jxl}. Since Higgs branch topological operators are annihilated by $Q_{1,2}^H$, they are placed on the fixed locus of $P_\tau$, namely the $S^1_\varphi$ at $\theta=\pi/2$ where $\tau$ shrinks, and can carry only Higgs branch R-symmetry.

An equivalent way to proceed is to consider the one-parameter family of supercharges $Q_{\beta}^H=Q_1^H+\beta Q_2^H$, $\beta$ being a phase, which satisfy the following algebra  
\begin{equation}\label{eq:algebraH}
\acomm*{Q_{\beta}^H}{\tilde{Q}_{\beta}^H}=P_\varphi+R_H+\mathbbm{i}rm
\end{equation}
where $\tilde{Q}_{\beta}^H$ is a suitable linear combination of the supercharges whose explicit expression is not relevant for the discussion, and $P_\varphi$ is the translation along the great circle $S^1_\varphi$. Here we include the central extension $m$ associated to the presence of the real mass term \cite{Dedushenko:2016jxl}. If we define the twisted translation $\hat P^H_\varphi=P_\varphi+R_H$, correlators of operators in the cohomology of $Q_{1,2}^H$ placed on $S^1_\varphi$ will be topological up to a mild $\varphi$-dependence $e^{-m r \varphi}$.

We now want to use this construction to define the topological sector in standard (non-conformal) ${\cal N} \geq 4$ SYM theories with vector multiplets coupled to hypermultiplets (see appendix \ref{appendix:conv} for conventions on the multiplets).
Since we are away the fixed points, we cannot rely on any argument based on representation theory of the superconformal algebra. Nevertheless, if $q_a$ and $\tilde{q}_a$ are the scalar fields of the hypermultiplet, a direct inspection shows that $q_1$ and $\tilde{q}_1$ are annihilated by $Q_{1,2}^H$ at $\theta=\pi/2$, $\varphi=0$. Acting with $\hat P^H_\varphi$, we obtain the corresponding twisted translated operators 
\begin{equation}\label{HBO}
Q(\varphi)=q_1(\varphi)\cos\frac{\varphi}{2}+q_2(\varphi)\sin\frac{\varphi}{2}\equiv u^a q_a, \quad \tilde Q(\varphi)=\tilde q_1(\varphi)\cos\frac{\varphi}{2}+\tilde q_2(\varphi)\sin\frac{\varphi}{2}\equiv u^a \tilde q_a\,,
\end{equation}
where $u^a=(\cos\varphi/2,\sin\varphi/2)$. Topological observables will be gauge invariant polynomials of $Q(\varphi)$ and $\tilde Q(\varphi)$. Correlators of twisted operators will depend on $\varphi$ only through $e^{-m r \varphi}$. 

\vskip 10pt
We can apply the same procedure to construct the topological sector of the Coulomb branch. In this case we introduce the two nilpotent supercharges
\begin{align}
Q_1^C&=\frac{1}{2}\left(Q_{11\dot2}+\mathbbm{i}Q_{12\dot2}+Q_{11\dot 1}+\mathbbm{i}Q_{12\dot1}+\frac{\mathbbm{i}}{2r}S_{11\dot2}-\frac{1}{2r}S_{12\dot2}-\frac{\mathbbm{i}}{2r}S_{11\dot 1}+\frac{1}{2r}S_{12\dot1}\right)\\ 
Q_2^C&=\frac{1}{2}\left(Q_{21\dot1}-\mathbbm{i}Q_{22\dot1}+Q_{21\dot2}-\mathbbm{i}Q_{22\dot2}+\frac{\mathbbm{i}}{2r}S_{21\dot1}+\frac{1}{2r}S_{22\dot1}-\frac{\mathbbm{i}}{2r}S_{21\dot2}-\frac{1}{2r}S_{22\dot2}\right) 
\end{align}
satisfying the following algebra
\begin{equation}
\acomm{Q_1^C}{Q_2^C}=\frac{4\mathbbm{i}}{r}\left(P_\tau+R_H+\mathbbm{i}r m\right)\,.
\end{equation}
Since this has the same structure of the algebra in \eqref{eq:algebraH} except for the replacement $\zeta \to m$, the rest of the construction works similarly to the Higgs branch. In particular, 
Coulomb branch operators in the $Q_{1,2}^C$ cohomologies are annihilated by $P_\tau$ and $R_H$ and live on $S^1_{\varphi}$ at $\theta = \pi/2$ where $\tau$ shrinks. 

As before, we can consider the family of cohomological supercharges $Q_\beta^C=Q_1^C+\beta Q_2^C$ satisfying
\begin{equation}
\acomm*{Q_\beta^C}{\tilde Q_\beta^C}=P_\varphi+R_C+\mathbbm{i}r\zeta
\end{equation}
It follows that operators in the $Q_\beta^C$-cohomology can be translated with the twisted translating generator $\hat{P}_\varphi=P_\varphi+R_C$ along $S^1_\varphi$ and they depend on $\varphi$ only through a factor $e^{- r\zeta \varphi}$. 

It is easy to realize that in the $Q_\beta^C$-cohomology there is only one operator built from local fields in the vector multiplet, which is 
\begin{equation}\label{eq:CBO}
\Phi(\varphi)=\Phi_{\dot{a}\dot{b}}v^{\dot{a}}v^{\dot{b}}\bigg|_{\theta=	\pi/2}\,, \quad v^{\dot a}=\frac{1}{\sqrt{2}}(e^{\mathbbm{i}\varphi/2}\,,e^{-\mathbbm{i}\varphi/2})
\end{equation} 
where $\Phi_{\dot{a}\dot{b}}$ are the triplet of dynamical scalars. Coulomb branch topological operators are given by gauge invariant polynomials of \eqref{eq:CBO}. We note that the cohomology contains also non-trivial monopole operators. However, they will not enter our derivation.

\section{An exact formula for integrated correlators}\label{sec:cohomological_equivalence_proof}

A localization scheme for the topological sector was proposed in \cite{Dedushenko:2016jxl, Dedushenko:2017avn, Dedushenko:2018icp} for ${\cal N} = 4$ SYM theories on $S^3$ coupled to hypermultiplets. However, the localization requires off-shell closed supersymmetry and cannot be straightforwardly applied to $\mathcal{N}\ge4$ Chern-Simons matter theory \cite{Gaiotto:2008sd, Hosomichi:2008jd}, including ABJM \cite{Aharony:2008ug, Aharony:2008gk, Hosomichi:2008jb}. For integrated correlators of dimension-one Higgs branch operators, an alternative formula was proposed in \cite{Agmon:2017xes, Binder:2019mpb}. There, it was conjectured that the 1d integrated correlators of Higgs branch operators are captured by the derivatives of the mass deformed partition function with respect to the mass parameters. For ABJM, this conjecture has been tested at weak coupling in \cite{Gorini:2020new}. 

We would like to prove the conjecture in general. The key idea is that Higgs branch operators sit in the bottom component of a conserved current multiplet. A mass deformation is realized, at the linear level, by coupling this multiplet to a background vector multiplet, and taking a rigid limit for the latter. The conjecture can then be restated as follows: the 1d integrated Higgs branch operators are $Q_\beta^H$-cohomologous to the mass deformation, meaning that the resulting generating functions coincide.

While the Higgs branch conjecture applies only to a specific class of operators, its validity is more general. It also should not depend on any specific Lagrangian realization of the theory. In the following, we will extend the formula for the generating function to Coulomb branch operators and propose a proof of the conjecture for both cases. We will also discuss the nonlinear coupling in the specific case of hypermultiplets.

\subsection{Coulomb branch operators}

We begin by discussing twisted Coulomb branch operators. The operator $\Phi_{\dot a \dot b}$ appearing in \eqref{eq:CBO} is the bottom component of the current multiplet associated with the $U(1)$ topological current $j_\mu\propto\epsilon_{\mu\nu\rho}F^{\nu\rho}$. This multiplet also contains the fermions $\lambda_{a\dot a}$ and the auxiliary fields $D_{ab}$ (see appendix \ref{appendix:back} for more details). For each $U(1)$ factor of the gauge group, there is a multiplet of this type. One can couple this multiplet to an abelian background twisted vector multiplet $\tilde{\mathcal{V}}_{\rm back}$. The relevant coupling is simply an $\mathcal{N}=4$ mixed Chern-Simons term. Taking the supersymmetry preserving rigid limit \eqref{FIback} for the background multiplet yields the Fayet-Iliopoulos (FI) term on $S^3$
\begin{equation}\label{FI}
S_{FI}=\mathbbm{i}\zeta\int_{S^{3}} d^3 x\sqrt{g}\left({h_a}^b{D_b}^a-\frac{1}{r}\bar{h}\indices{^{\dot{a}}_{\dot{b}}} \Phi\indices{^{\dot{b}}_{\dot{a}}}\right)\,.
\end{equation}
We now show how to relate the integrated correlators of topological Coulomb branch operators of the type appearing in eq. \eqref{eq:CBO}, and derivatives of this term w.r.t. to the FI parameter $\zeta$. 

We can infer the precise formula for the equality of generating functions by building on the localization result of \cite{Dedushenko:2016jxl}. After localization, the effect of the FI term is the multiplication of the matrix integrand function by a term $\exp\left(-8\pi^2\mathbbm{i} r\zeta\sigma \right)$, where $\sigma$ is the integration variable. 
One may also express $\sigma$ in terms of the twisted field $\Phi(\varphi)$ defined in eq. \eqref{eq:CBO} integrated over $S_\varphi^1$.
Therefore, following localization, derivatives of the FI deformed partition function compute the simpler 1d integral of twisted Coulomb branch operators\footnote{The set of Coulomb branch operators also contains monopole operators. In principle, these can appear in our integrated correlators. However, this is not the case, as can be shown using an argument based on the topological symmetry. This symmetry is just a shift invariance for the dual photon $\gamma$. Therefore, only derivatives of $\gamma$ can appear in any effective Lagrangian.  Since the FI-term is a mass term, no coupling to $\gamma$ is included in the deformation. Therefore, the dual photon does not appear at all in the FI-term and monopole operators do not appear in our formulae.} via the following formula 
\begin{equation}\label{cbo:equiv}
\langle \int_{S^1_{\varphi_1}}d\varphi_1\,\Phi(\varphi_1)\cdots\int_{S^1_{\varphi_n}}d\varphi_n\,\Phi(\varphi_n)\rangle=\frac{1}{(4\pi \mathbbm{i}r^2)^n}\frac{\partial^n}{\partial \zeta^n}Z[\zeta]\,,
\end{equation}
where $Z[\zeta]$ is the FI-deformed partition function.
The equality can be argued on the basis of a Ward identity, which holds in any 3d $\mathcal{N}=4$ theory, regardless of any localization procedures or Lagrangian realization of the theory. 
The Ward identity is the result of the following cohomological equivalence
\begin{equation}\label{Ward:coul}
S_{FI}=-4\pi\mathbbm{i}r^2\int_{S^1_\varphi}d\varphi\,\Phi_{\dot{a}\dot{b}}v^{\dot{a}}v^{\dot{b}}+\delta_\xi\left[ \int_{S^3} d^3x \sqrt{g}\,\tilde\eta^{a\dot a}\lambda_{a\dot a}\right]\,.
\end{equation}
where $\delta_\xi$ stands for a variation generated by $Q_\beta^C$ with parameter $\xi_{a \dot a}$. Let us assume that the last term on the rhs is also $\delta_\xi$ closed. Then the difference between the expressions leading, after taking expectation values, to \eqref{cbo:equiv} is $\delta_\xi$ exact, while preserving supersymmetry. It is well known that such expressions vanish after taking expectation values, thereby establishing \eqref{cbo:equiv}. In the following, we will show that a spinor $\tilde\eta_{a\dot a}$ exists such that \eqref{Ward:coul} holds.

The presence of both 3d and 1d terms in \eqref{Ward:coul} may be surprising. However, a \emph{mild singularity} for $\tilde{\eta}_{a\dot{a}}$ at $\theta=\pi/2$ can yield such a contribution.
Since $\delta_\xi\lambda_{a\dot a} = \mathbbm{i}\gamma^\mu{\xi_a}^{\dot{b}}\nabla _\mu\Phi_{\dot{b}\dot{a}} + \cdots$ (see eq. \eqref{lambdavar}) involves a derivative, integration by parts can give rise to a boundary term expressed as a delta function at $\theta=\pi/2$
\begin{equation}\label{delta:coulomb}
 -\frac{2\mathbbm{i}}{r}\int_{S^3}d^3x\sqrt{g}\,\delta\left(\theta-\pi/2\right)\Phi_{\dot{a}\dot{b}}v^{\dot{a}}v^{\dot{b}} \subset\delta_\xi\left[ \int_{S^3} d^3x \sqrt{g}\,\tilde\eta^{a\dot a}\lambda_{a\dot a}\right]\,.
\end{equation}
In this way, cohomological equivalence reduces the 3d functional to a 1d one, in the spirit of \cite{Zaboronsky:1996qn, Pestun:2009nn, Dedushenko:2016jxl, Mezei:2018url, Wang:2020seq}.

Let us thus discuss the boundary contribution in detail. Integrating by parts the expression $\tilde{\eta}^{a\dot{a}} \delta_{\xi} \lambda_{a\dot a} \sim \tilde{\eta}^{a\dot{a}}\gamma^\mu{\xi_a}^{\dot{b}}\nabla _\mu\Phi_{\dot{b}\dot{a}}$, we obtain the following total derivative
\begin{equation}
-\mathbbm{i}\int_{S^3}\left[\sqrt{g}\,\nabla_{\mu}\left(w^\mu_{\dot a\dot{b}}\Phi^{\dot a\dot b}\right)\right]\,,
\end{equation}
where 
\begin{equation}
w^\mu_{\dot a\dot{b}} \equiv {\tilde{\eta}^a}_{\; \; \dot a} \gamma^\mu \xi_{ a\dot b}\,.
\end{equation}
Let us assume that $w^\mu_{\dot a\dot{b}}$ has a singularity at $\theta=\pi/2$ such that $w^\theta_{\dot a\dot{b}}$ has a pole $\left(\theta-\pi/2\right)^{-1}$.
We excise a small solid torus $T$, which can be taken to be $\partial_{\varphi}$ and $\partial_{\tau}$ invariant, around the circle at $\theta=\pi/2$. The size of the torus is determined by a cutoff $\theta_{0}$, such that the torus extends along $\theta\in\left(\theta_{0},\pi/2\right],\varphi\in\left[0,2\pi\right),\tau\in\left[0,2\pi\right)$.
In the limit where the size of the torus shrinks to zero, the boundary contribution can be written as
\begin{align}
-\mathbbm{i}\int_{S^3}\left[\sqrt{g}\,\nabla_{\theta}\left(w^\theta_{\dot a\dot{b}}\Phi^{\dot a\dot b}\right)\right]  =
- 2\pi\mathbbm{i} \, r^3 \oint_{\varphi}\left[\Phi^{\dot a\dot b}\Big|_{\theta=\pi/2}\,\lim_{\theta_{0}\rightarrow\pi/2}\left(\left(\pi/2-\theta_{0}\right)w^\theta_{\dot a\dot{b}}\Big|_{\theta=\theta_{0}}\right)\right],
\end{align}
where we have used the fact that $\Phi^{\dot a\dot b}$ is non-singular. The factor of $2\pi$ comes from integrating over $\tau$, whereas $r^3$ comes from $\sqrt{g}$. In order for the above to reproduce the delta function term in \eqref{delta:coulomb}, we need to require
\begin{equation}\label{c0}
w^\theta_{\dot a\dot b}\sim \frac{2}{r} \, \frac{v_{\dot a} v_{\dot b}}{\theta - \pi/2}\,.
\end{equation}
In addition, the remaining terms from $\int_{S^3} d^3x \sqrt{g} \, \tilde{\eta}^{a\dot a} \delta_\xi \lambda_{a \dot a}$  must reproduce the contributions in $S_{FI}$ proportional respectively to $D_{ab}$ and $\Phi_{\dot a\dot b}$, implying the following set of equations
\begin{align}
w_\mu^{\dot a\dot b}\,\epsilon_{\dot a\dot b}&=0\,,\label{c1}\\
\tilde\eta\indices{_{(a}^{\dot b}}\xi_{b)\dot b}-\mathbbm{i} h_{ab}&=0\,,\label{c2}\\
\nabla_\mu w^\mu_{\dot a\dot b}-2{\tilde\eta^a}_{(\dot a}\xi'_{|a|\dot b)}+\frac{1}{r}\bar h_{\dot a\dot b}&=0\label{c3}\,.
\end{align}
Finally, we must demand that the $\delta_\xi$-exact term is itself invariant under the action of $\delta_\xi$.

The set of equations (\ref{c0}-\ref{c3})  is an overconstrained system of nine linear algebraic equations and three linear differential equations for the eight unknowns in 
$\tilde{\eta}_{a\dot{a}}$, further constrained by the demand that $\tilde{\eta}_{a\dot{a}}$ is annihilated by $\delta^2_\xi$. 
Nevertheless, one can show that a solution exists.\footnote{The solution space is actually three dimensional, with any choice of solution being singular at $\theta=\pi/2$. The explicit form for the solutions is very complicated and not extremely enlightening, so we avoid writing them explicitly.} We conclude that the cohomological equivalence in eq. \eqref{Ward:coul} holds.

\subsection{Higgs branch operators}

We now move to twisted Higgs branch operators. Dimension-one scalar operators can be seen as the bottom components of $\mathcal{N}=4$ linear multiplets $\Sigma^A =(J^A_{ab},\, \chi^A_{a\dot{a}}, \, j^A_\mu, \, K^A_{\dot{a}\dot{b}})$.\footnote{Here $A$ is an index which runs from 1 to the rank of the flavor symmetry Lie algebra.}
The dimension-one scalars $J^A_{ab}$ are in the $(\mathbf{3},\mathbf{1})$ of the R-symmetry group, $\chi^A_{a\dot{a}}$ are the fermion partners of dimension $3/2$ in the $(\mathbf{2},\mathbf{2})$ of the R-symmetry group, $j_\mu^A$ are the flavor conserved currents, and $ K^A_{\dot{a}\dot{b}}$ are dimension-two scalars in the $(\mathbf{1},\mathbf{3})$ of the R-symmetry group. 

We couple $\Sigma^A$ to a background vector multiplet $\mathcal{V}^A_{\textup{back}}$ fixed to a rigid supersymmetric configuration, to produce the real mass deformation. This amounts to modifying the action by the following terms
\begin{equation}\label{massdef}
S_{\textup{mass}}=m^A \int_{S^3} d^3x\sqrt{g}\left(-\mathbbm{i}h^{ab}J^A_{ab}+\bar h^{\dot a\dot{b}}K^A_{\dot{a}\dot{b}}\right) +O\left(m^2\right) \,.  
\end{equation}
The terms of order $m^2$ are additional terms needed to preserve supersymmetry.

According to \cite{Dedushenko:2016jxl}, for theories without Chern-Simons terms localization on $S^3$ shows that Higgs branch operators are captured by a 1d theory with a quadratic action $S_\sigma[Q\,,\tilde{Q}]$, coupled to the standard matrix model.
The 1d fundamental degrees of freedom $Q,\,\tilde{Q}$ are the twisted operators defined in eq. \eqref{HBO}.  We can introduce a mass deformation at the level of the 1d theory by adding the following term in the 1d action
\begin{equation}\label{HBO:loc}
S_\sigma[Q\,,\tilde{Q}]\to S_\sigma[Q\,,\tilde{Q}]- 4\pi r^2 m^A \int_{-\pi}^\pi d\varphi \,\tilde{Q}(\varphi)T^A Q(\varphi)\,.
\end{equation}
where $T^A$ are the generators of the flavor symmetry in the proper representation. The operators $J^A(\varphi)\equiv - \tilde{Q}(\varphi)T^A Q(\varphi)$ are the twisted operators $J^A(\varphi) = J^A_{ab}u^au^b$. 
Localization can be used to show that the following equality holds
\begin{equation}\label{conjecture}
\hspace{-0.3cm} \Big\langle \int d\varphi_1\, J^{A_1}(\varphi_1)\dots \int d\varphi_n \,J^{A_n}(\varphi_n)\Big\rangle=\left( - \frac{1}{4\pi r^2} \right)^{\! n} \, \frac{1}{Z}\frac{\partial^n Z[m_1,\dots ,m_n]}{\partial m^{A_1}\dots \partial m^{A_n}}\bigg|_{m^{A_1}\!, m^{A_2}\!, \cdots = 0}
\end{equation}
It has been conjectured that this identity should be true even in theories where the localization argument cannot be carried though. We now show that this is the case, relying on a cohomological argument, in analogy with the Coulomb branch case. 

We introduce a generating function for a $\mathfrak{u}(1)$ flavor current
\begin{equation}\label{eq:massaction}
Z_{\textup{flavor}}[m]=\Big\langle \exp\left(m\int_{S^3}d^3x\sqrt{g}\left(-\frac{\mathbbm{i}}{r}h^{ab}J_{ab}+\bar h^{\dot a\dot b}K_{\dot a \dot b} \right)+O(m^2)\right)\Big\rangle\,,
\end{equation}
and, in analogy with eq. \eqref{HBO:loc}, a 1d generating function for integrated Higgs branch operators
\begin{equation}
Z_{\textup{J}}[m]=\Big\langle \exp\left( -4\pi r^2 m\oint_{S^1_{\varphi}}    J(\varphi)   \right)\Big\rangle\,.
\end{equation}
The cohomological argument states that mass derivatives of $Z_{\textup{flavor}}[m]$ and $Z_{\textup{J}}[m]$ are in the same $Q_\beta^H$ cohomology class, that is 
\begin{equation}\label{Higgs:fundeq}
\pdv{m}\left(S_{\textup{mass}}[m]-4\pi r^2 m\oint_{S^1_{\varphi}}    J(\varphi)   \right)= \acomm{Q_\beta^H}{\dots}\,,
\end{equation}
where $S_{\textup{mass}}[m]$ is the action appearing in the exponent of \eqref{eq:massaction}. Establishing this identity is sufficient in order to show that eq. \eqref{conjecture} holds.

At linear order, showing that identity \eqref{Higgs:fundeq} holds amounts to showing that a spinor $\tilde{\xi}_{a\dot{a}}$ exists such that
\begin{equation}
\delta_{\xi}\int_{S^3}\left[\sqrt{g}\tilde{\xi}_{a\dot{a}}\chi^{a\dot{a}}\right]=\int_{S^3}\sqrt{g}\left[-\frac{\mathbbm{i}}{r}h^{ab}J_{ab}+\bar{h}^{\dot{a}\dot{b}}K_{\dot{a}\dot{b}}+2r^{2}\frac{1}{\sqrt{g}}\delta\left(\theta-\pi/2\right)J_{ab}u^{a}u^{b}\right]\,,\label{eq:cohomological_equivalence}
\end{equation}
where $\delta_\xi$ indicates the variation generated by $Q_\beta^H$ with parameter $\xi_{a \dot a}$. The derivation goes along the same lines as that of the Coulomb branch operators. Again, we require the appearance of a singularity in $\tilde\xi_{a\dot a}$ in order to obtain a delta function localizing at $\theta=\pi/2$. 
Defining ${v^{\mu}}\indices{_{ab}}\equiv \tilde{\xi}\indices{_a^{\dot b}}\gamma^\mu\xi\indices{_{b\dot b}}$, we need to require the following asymptotic behavior 
\begin{equation}\label{asym}
{v^{\theta}}^{(ab)}\sim2r^{2}\left(\pi/2-\theta\right)^{-1}u^{a}u^{b}\,.
\end{equation}
This condition must be supplemented by the following conditions
\begin{align}
{{v^{\mu}}_{a}}^{a}&=0\,,\\
\tilde{\xi}^{a(\dot{a}}{\xi_a}^{\dot{b})}&=-\bar{h}^{\dot{a}\dot{b}}\,,\label{cK}\\
\nabla_{\mu}{v^{\mu}}^{(ab)}-2{\tilde{\xi}}^{(a|\dot{a}|}{\xi'^{b)}}_{\dot{a}}&=-\frac{\mathbbm{i}}{r}h^{ab}\,,
\end{align}
which arise requiring that all the linear terms in $S_{\textup{mass}}[m]$ get correctly reproduced.
Indeed, we find a solution. This is parametrized by three regular function of $\theta$ and $\varphi$: $h_1$, $h_2$, and $h_3$. Its explicit form reads
\begin{align}
\tilde{\xi}_{1,1\dot1}&=e^{-\mathbbm{i} \tau } h_1(\theta ,\varphi )\\
\tilde{\xi}_{2,1\dot1}&=\frac{1}{\beta }\left(h_2(\theta ,\varphi )+\frac{1}{\sqrt{\sin \theta  \cos \varphi +1}}\right)\notag\\
\tilde{\xi}_{1,1\dot2}&=h_2(\theta ,\varphi )\notag\\
\tilde{\xi}_{2,1\dot2}&=\frac{e^{\mathbbm{i} \tau } \tan \theta  \sin \varphi }{\sqrt{\sin \theta  \cos \varphi +1}}-\beta  e^{\mathbbm{i} \tau } h_1(\theta ,\varphi )\notag\\
\tilde{\xi}_{1,2\dot1}&=e^{-\mathbbm{i} \tau } h_3(\theta ,\varphi )\notag\\
\tilde{\xi}_{2,2\dot1}&=-e^{\mathbbm{i} \tau } \sec \theta  \left(e^{-\mathbbm{i} \tau } \sin \theta  \sin \varphi  \,\,h_3(\theta ,\varphi )+e^{-\mathbbm{i} \tau } h_1(\theta ,\varphi ) (\sin \theta  \cos \varphi +1)\right)\notag\\
\tilde{\xi}_{1,2\dot2}&=\beta  \left(-e^{\mathbbm{i} \tau }\right) \sec \theta  \left(e^{-\mathbbm{i} \tau } \sin \theta  \sin \varphi \, h_3(\theta ,\varphi )+e^{-\mathbbm{i} \tau } h_1(\theta ,\varphi ) (\sin \theta  \cos \varphi +1)\right)\notag\\
\tilde{\xi}_{2,2\dot2}&=-e^{\mathbbm{i} \tau } \left(\beta  h_3(\theta ,\varphi )+\sec (\theta ) \sqrt{\sin (\theta ) \cos (\varphi )+1}\right)\notag\,.
\end{align}
This solution is also invariant under the action of $\delta^2_\xi$. Therefore, we conclude that eq. \eqref{Higgs:fundeq} is verified at linear level.

Unlike the Coulomb branch operators, the Higgs branch coupling involves quadratic terms in $m$. In the next section, we show that the cohomological equivalence persists beyond the linear order, at least in the simple example of the current multiplet built from a hypermultiplet. Presumably, one may argue that this is always true, based on the supersymmetry and gauge symmetry preserving nature of the complete non-linear coupling (see e.g. \cite{Closset:2013vra}).

\subsection{Beyond the linearized analysis: an explicit example}

For a generic theory we consider the current multiplet built from a hypermultiplet, coupled to an abelian background vector multiplet of the form \eqref{mback}. 

The components of the current multiplet are worked out in appendix \ref{App:current}. In terms of the hypermultiplet components, choosing the background \eqref{mback} they are given by 
\begin{align}
& J_{ab}= - \tilde q_{(a}q_{b)}, \qquad  \qquad \qquad \quad \; \, \chi_{a\dot a}=-\mathbbm{i}\left(\tilde{q}_a\psi_{\dot a}+\tilde{\psi}_{\dot a}q_a\right) \notag\\
& K_{\dot a\dot b}=\mathbbm{i}\tilde\psi_{(\dot a}\psi_{\dot b)}-m \bar{h}_{\dot a \dot b} \, \tilde{q}^b q_b\, \qquad 
j_\mu =\mathbbm{i}\tilde{q}^a\nabla_\mu q_a-\mathbbm{i}\nabla_\mu\tilde{q}^a\,q_a-\tilde{\psi}^{\dot b}\gamma_\mu\psi_{\dot b}
\end{align}
We note that some of the components depend on the background fields. In particular, we are interested in the $m$ dependence of $K_{\dot a\dot b}$, thus we rewrite
\begin{equation}\label{splitting}
K_{\dot a\dot b}=K^{\textup{lin}}_{\dot a\dot b}- m \bar{h}_{\dot a \dot b} \, \tilde{q}^b q_b\,,
\end{equation}
where $K^{\textup{lin}}_{\dot a\dot b}\equiv \mathbbm{i}\tilde\psi_{(\dot a}\psi_{\dot b)}$ is the $m=0$ piece. 

The hypermultiplet action can be read from \eqref{Shyper} by substituting the \eqref{mback} background
\begin{align}\label{Shyper2}
S_{\textup{hyper}}[m]&=\int d^3x\sqrt{g}\Bigg[ \nabla^\mu\tilde{q}^a\nabla _\mu q_a - \mathbbm{i}\tilde{\psi}^{\dot{a}}\gamma^\mu\nabla_\mu\psi_{\dot{a}}+\frac{3}{4r^2}\tilde{q}^a q_a \notag \\
& \qquad \qquad \qquad -\mathbbm{i}\frac{m}{r} \tilde{q}^a{h_a}^b q_b -\mathbbm{i}m \tilde{\psi}^{\dot{a}}\, \bar{h}_{\dot{a}}^{\; \; \dot{b}} \, \psi_{\dot{b}} - \frac{m^2}{2}\tilde{q}^a \, \bar{h}^{\dot{a}\dot{b}}\bar{h}_{\dot{a} \dot{b}} \, q_a \Bigg] \notag \\
& \equiv S_{\textup{hyper}}[0]+m S_{\textup{lin}}+\frac{1}{2}m^2 \frac{d^2S}{dm^2}  \,. 
\end{align}
where we have defined 
\begin{align}\label{quadratic}
&S_{\rm lin} = - \mathbbm{i} \int d^3x \sqrt{g} \, (\frac{1}{r} \tilde{q}^a{h_a}^b q_b + \tilde{\psi}^{\dot{a}}\, \bar{h}_{\dot{a}}^{\; \; \dot{b}} \, \psi_{\dot{b}}) = 
\int d^3x \sqrt{g} \, ( - \frac{\mathbbm{i}}{r} h^{ab} J_{ab} + \bar{h}^{\dot a \dot b} K_{\dot a \dot b}^{\rm lin} )  \notag \\ 
& \frac{d^2 S}{dm^2} = -\int d^3x \sqrt{g} \, \tilde{q}^a \, \bar{h}^{\dot{a}\dot{b}}\bar{h}_{\dot{a} \dot{b}} \, q_a \, .
\end{align}
Here, we are interested in the full non-linear coupling. 

The $m$-dependent term in $K_{\dot a\dot b}$, eq. \eqref{splitting}, enters the variation of $\chi_{a\dot a}$ (see eq. \eqref{susy:curr2})  and consequently affects the cohomological equivalence \eqref{eq:cohomological_equivalence} by one extra term linear in $m$. Here we prove that this extra piece reproduces exactly the quadratic coupling in the 3d action \eqref{Shyper2}.  

To this end, we observe that retaining the non-linear terms in $m$, the cohomological equivalence \eqref{Higgs:fundeq} is equivalent to 
\begin{equation}\label{variation2}
S_{\textup{lin}}+m \frac{d^2S}{dm^2}-4\pi r^2 \oint_{S^1_{\varphi}}    J(\varphi) = \delta\left[\int d^3x\sqrt{g}\,\tilde{\xi}^{a\dot{a}}\,\chi_{a\dot{a}}\right]\,.
\end{equation}
where $S_{\textup{lin}}$ and $\tfrac{d^2 S}{dm^2}$ are given in \eqref{quadratic}.  
The equivalence at linear order has been already established in the previous section. We then focus only on terms in \eqref{variation2} proportional to $m$. Using variations \eqref{susy:curr2}, the splitting in \eqref{splitting} and eq. \eqref{cK} it is easy to see that 
\begin{align}
- \int d^3x\sqrt{g}\,\tilde{\xi}^{a\dot{a}}\, \delta \chi_{a\dot{a}} & \to - \int d^3x\sqrt{g}\,\tilde{\xi}^{a\dot{a}}\, \xi\indices{_a^{\dot b}}K_{\dot a\dot b} \notag \\
& \to m \int d^3x\sqrt{g}\,\tilde{\xi}^{a\dot{a}}\, \xi\indices{_a^{\dot b}}\bar{h}_{\dot a \dot b} \, \tilde{q}^b q_b = -m \int d^3x\sqrt{g}\,\bar{h}^{\dot a \dot b} \bar{h}_{\dot a \dot b} \, \tilde{q}^b q_b
\end{align}
and this coincides with $ \frac{d^2S}{dm^2}$ in \eqref{quadratic}. We have thus shown that at least in this case the cohomological equivalence persists beyond the linear order. 

While the linear coupling is universal and does not depend on the specific theory, the second order term might depend on details of the theory, such as the off-shell closure of the supercharge used for the cohomological equivalence. For this, reason the computation does not apply to all possible theories. However, we believe that supersymmetry and gauge invariance control the non-linear terms, and that a mechanism similar to the one discussed in this example will ensure the validity of \eqref{Higgs:fundeq} regardless of the specific theory.

\section{Conclusions}\label{sec:conclusions}

We have provided a proof of the conjecture presented in \cite{Agmon:2017xes, Binder:2019mpb}, stating that the generating function of correlators of integrated Higgs and Coulomb branch operators coincides with the mass and FI deformed three sphere partition function. The result applies to any 3d $\mathcal{N}\ge 4$ theory. Specifically, the theory needs not have off-shell closed supersymmetry, or even a Lagrangian description. The technique used to establish the result, a cohomological equivalence between a deformation of the theory which is integrated over the entire spacetime and one which is restricted to a submanifold, is itself interesting. We believe it may have many additional applications, for instance to theories with defects (for a recent review, see \cite{Penati:2021tfj}).

\section*{Acknowledgments}

We would like to thank Luca Griguolo and Domenico Seminara for interesting discussions. This work was supported in part by Italian Ministero dell'Istruzione, Universit\`a e Ricerca (MUR), and Istituto Nazionale di Fisica Nucleare (INFN) through the ``Gauge and String Theory'' (GAST) and the ``Gauge theories, Strings, Supergravity'' (GSS) projects. The work of IY was financially supported by the European Union's Horizon 2020 research and innovation programme under the Marie Sklodowska-Curie grant agreement No. 754496 - FELLINI.  

\appendix

\section{Conventions}\label{appendix:conv}

We follow the conventions of \cite{Dedushenko:2016jxl}. Here, we just recall what is needed for our derivation. We work on $S^3$, which is defined by the following embedding in $\mathbb{R}^4$
\begin{equation}
X_1+\mathbbm{i}X_2=r\,e^{\mathbbm{i}\tau}\cos\theta\,,\qquad X_3+\mathbbm{i}X_4=r\,e^{\mathbbm{i}\varphi}\sin\theta \,.
\end{equation}
with the toroidal coordinates valued as, $\theta\in[0,\,\frac{\pi}{2}]$, $\varphi\in[0,\,2\pi)$, $\tau\in[0,\,2\pi)$. The induced metric is
\begin{equation}\label{eq:metric}
ds^2=r^2\left(d\theta^2+\cos^2\theta\, d\varphi^2+\sin^2\theta \,d\tau^2\right) \,.
\end{equation}

\vskip 10pt
A generic ${\cal N} \geq 4$ gauge theory on $S^3$ is described in terms of vector multiplets and hypermultiplets.  
Field components are labeled by a Lorentz spin, and gauge group $G$ and R-symmetry $\mathfrak{su}(2)_H\oplus\mathfrak{su}(2)_C$ representation indices. We denote Lorentz spinor indices with $\alpha=1,2$, while $a,\dot{a}=1,2$ indicate $\mathfrak{su}(2)_H$ and $\mathfrak{su}(2)_C$ indices, respectively.  

The $\mathcal{N}=4$ vector multiplet, $\mathcal{V}=\left(A_\mu, \, \lambda_{a\dot{a}}, \, \Phi_{\dot{a}\dot{b}}, \, D_{ab}\right)$, contains 
the gauge field $A_\mu$, the gauginos $\lambda_{\alpha,a\dot{a}}$, the dimension-one scalar fields $\Phi_{\dot{a}\dot{b}}$ and the dimension-two scalar fields $D_{ab}$, transforming in the $(\mathbf{1},\mathbf{1})$, $(\mathbf{2},\mathbf{2})$, $(\mathbf{1},\mathbf{3})$ and $(\mathbf{3},\mathbf{1})$ of $\mathfrak{su}(2)_H\oplus\mathfrak{su}(2)_C$, respectively. For our purposes we need to recall the SUSY variation of the (abelian) gaugino
\begin{align}\label{lambdavar}
\delta_\xi\lambda_{a\dot{a}}&=-\frac{\mathbbm{i}}{2}\epsilon^{\mu\nu\rho}\gamma_{\rho} \, \xi_{a\dot{a}}F_{\mu\nu}-{D_{a}}^b \, \xi_{b\dot{a}}-\mathbbm{i}\gamma^\mu{\xi_a}^{\dot{b}}\nabla _\mu\Phi_{\dot{b}\dot{a}}+2\mathbbm{i}{\Phi_{\dot{a}}}^{\dot{b}} \, \xi'_{a\dot{b}} \,,  
\end{align}
where $\xi_{a\dot{a}}, \xi'_{a\dot{a}}$ are Killing spinors solving the equations $\nabla_\mu \xi_{a \dot a} = \gamma_\mu \xi'_{a \dot a}$, $\nabla_\mu \xi'_{a \dot a} = - \tfrac{1}{4r^2} \gamma_\mu \xi_{a \dot a}$ on $S^3$. 

Similarly, a twisted vector multiplet $\tilde{\mathcal{V}}= \left(\tilde{A}_\mu, \, \tilde{\lambda}_{\alpha, a\dot{a}}, \, \tilde{\Phi}_{ab}, \,\tilde{D}_{\dot{a}\dot{b}}\right)$ can be constructed, whose components have the same Lorentz structure as the previous ones, but differ for the $\mathfrak{su}(2)_H\oplus\mathfrak{su}(2)_C$ representations that are now $(\mathbf{1},\mathbf{1})$, $(\mathbf{2},\mathbf{2})$, $(\mathbf{3},\mathbf{1})$ and $(\mathbf{1},\mathbf{3})$.
The SUSY variation for the twisted gaugginos in the abelian case is
\begin{align}
\delta_\xi\tilde\lambda_{a\dot{b}}&=-\frac{\mathbbm{i}}{2}\epsilon^{\mu\nu\rho}\gamma_{\rho}\xi_{a\dot{b}}\tilde F_{\mu\nu}- D\indices{_{\dot b}^{\dot c}}\xi_{a\dot{c}}-\mathbbm{i}\gamma^\mu{\xi_b}^{\dot{b}}\nabla _\mu\tilde\Phi_{ab}+2\mathbbm{i}{\tilde\Phi_{a}}^{b}\xi'_{b\dot{b}}\,.
\end{align}

The hypermultiplet $\mathcal{H}$ belongs to a unitary representation $\mathcal{R}$ of the gauge group $G$. Its field components are $\mathcal{H}=\left( q_a, \tilde{q}^a, \psi_{\dot{a}},\tilde{\psi}_{\dot{a}} \right)$, where $q_a$ and $\psi_{\alpha, \dot{a}}$ are scalar and fermion fields transforming in the $\mathcal{R}$ of $G$, and in the $(\mathbf{2}, \mathbf{1})$ and $(\mathbf{1},\mathbf{2})$ of $\mathfrak{su}(2)_H\oplus\mathfrak{su}(2)_C$, respectively. Similarly, $\tilde{q}^a$  and $\tilde{\psi}^{\dot{a}}$ are scalar and fermion fields transforming in the $\bar{\mathcal{R}}$ of $G$, and in the $({\mathbf{2}},\mathbf{1})$ and $(\mathbf{1}, {\mathbf{2}})$ of the R-symmetry group. The SUSY transformations read  
\begin{subequations}\label{hyper:var}
\begin{align}
\delta_\xi q^a&=\xi^{a\dot{b}}\psi_{\dot{b}}\,,  &\delta_\xi\psi_{\dot{a}}&=\mathbbm{i}\gamma_\mu\xi_{a\dot{a}}\mathcal{D} _\mu q^a+\mathbbm{i}\xi'_{a\dot{a}}q^a-\mathbbm{i}\xi_{a\dot{c}}\Phi\indices{^{\dot{c}}_{\dot{a}}}q^a \,, \\
\delta_\xi\tilde{q}^a&=\xi^{a\dot{b}}\tilde{\psi}_{\dot{b}}\,,         &\delta_\xi \tilde{\psi}_{\dot{a}}&=\mathbbm{i}\gamma_\mu\xi_{a\dot{a}}\mathcal{D} _\mu \tilde{q}^a+\mathbbm{i}\xi'_{a\dot{a}}\tilde{q}^a+\mathbbm{i}\xi_{a\dot{c}}\Phi\indices{^{\dot{c}}_{\dot{a}}}\tilde{q}^a  \,,
\end{align}
\end{subequations}
where $\mathcal{D}_\mu = \nabla_\mu - iA_\mu$.

The action for a hypermultiplet minimally coupled to a gauge multiple reads
\begin{align}\label{Shyper}
S_{\textup{hyper}}&=\int d^3x\sqrt{g}\Bigg[ \mathcal{D}^\mu\tilde{q}^a\mathcal{D} _\mu q_a - \mathbbm{i}\tilde{\psi}^{\dot{a}}\gamma^\mu\mathcal{D}_\mu\psi_{\dot{a}}+\frac{3}{4r^2}\tilde{q}^a q_a+\mathbbm{i}\tilde{q}^a{D_a}^b q_b-
\frac{1}{2}\tilde{q}^a\Phi^{\dot{a}\dot{b}}\Phi_{\dot{a}\dot{b}} q_a+ \notag\\
&-\mathbbm{i}\tilde{\psi}^{\dot{a}}{\Phi_{\dot{a}}}^{\dot{b}}\psi_{\dot{b}}+\mathbbm{i}\left(\tilde{q}^a{\lambda_a}^{\dot{b}}\psi_{\dot{b}}+\tilde{\psi}^{\dot{a}}{\lambda^{b}}_{\dot{a}} q_b\right)\Bigg] \,.
\end{align}

\section{Background flavor symmetry}\label{appendix:back}

In this appendix we review some relevant couplings to background flavor symmetries.

A current multiplet $\Sigma = (J_{ab},\, \chi_{a\dot{a}}, \,  j_\mu \, , K_{\dot{a}\dot{b}})$ couples to a background vector field $\mathcal{V}_{\rm back} = \left(A_\mu, \, \lambda_{a\dot{a}}, \, \Phi_{\dot{a}\dot{b}}, \, D_{ab}\right)$ as
\begin{equation}\label{coupl}
\int_{S^3} d^3x\sqrt{g}\left(A_\mu j^\mu +\mathbbm{i}D^{ab}J_{ab}+\Phi^{\dot a\dot{b}}K_{\dot{a}\dot{b}}+\lambda^{a\dot{a}}\chi_{a\dot{a}}+O\left(\mathcal{V}^2 \right) \right)\,.
\end{equation}
Imposing supersymmetry invariance of this action and exploiting the SUSY transformations of the vector components \cite{Dedushenko:2016jxl}, we can read the linearized variation of the $\chi_{a\dot a}$ fermion 
\begin{equation}
\delta_\xi\chi_{a\dot{a}}=\frac{\mathbbm{i}}{2}\gamma_{\mu}\xi_{a\dot{a}}j^{\mu}+2{\xi^{'b}}_{\dot{a}}J_{ab}+\gamma^{\mu}{\xi^{b}}_{\dot{a}}\nabla_{\mu}J_{ab}+{\xi_{a}}^{\dot{b}}K_{\dot{a}\dot{b}}.
\end{equation}
A rigid supersymmetry preserving background is obtained imposing $\lambda_{a\dot a}=0$ and $\delta\lambda_{a\dot a}=0$. We take the following solution 
\begin{equation}\label{mback}
\Phi_{\dot{a}\dot{b}}=m \,\bar h_{\dot a\dot b}\,, \quad D_{ab}=-\frac{m}{r} h_{a b}\,, \quad A_\mu=0\,,\quad \lambda_{a\dot a}=0\,. 
\end{equation}
where the parameter $m$ is the real mass. If we restore the adjoint index $A$, the background reproduces the real mass deformation of eq. \eqref{massdef}.

A similar construction can be applied to obtain the FI-term \eqref{FI}. Because of the Bianchi identities, for each $U(1)$ factor of the gauge group there is an associated conserved current $j_\mu \sim \epsilon_{\mu\nu\rho}F^{\nu\rho}$. We will refer to this symmetry as the topological symmetry. In supersymmetric theories, the topological current sits in a linear multiplet built from vector multiplet components. Such a multiplet can be coupled to an abelian twisted vector multiplet background $\tilde{\mathcal{V}}_{\rm back} = \left(\tilde{A}_\mu, \, \tilde{\lambda}_{\alpha, a\dot{a}}, \, \tilde{\Phi}_{ab}, \,\tilde{D}_{\dot{a}\dot{b}}\right)$. 
The coupling, originally found in \cite{Brooks:1994nn}, can be derived by constructing the supersymmetric completion of the BF term. We obtain
\begin{equation} 
S_{\textup{SBF}}=\frac{\mathbbm{i}}{4\pi}\int_{S^3} d^3x \sqrt{g} \,\left(\epsilon^{\mu\nu\rho} F_{\mu\nu} \tilde A_\rho-\tilde\lambda^{a\dot{a}}\lambda_{a\dot a}-\tilde\Phi^{ab}D_{ab}-\Phi^{\dot{a}\dot{b}} \tilde{D}_{\dot{a}\dot{b}}\right)\,.
\end{equation}
This is a topological action which is invariant under the full superconformal algebra. However, the choice of the background $\tilde{\lambda}_{a\dot a}=0, \delta\tilde{\lambda}_{a\dot a}=0$ breaks conformal invariance and selects the Poincaré subalgebra. The explicit rigid background that reproduces \eqref{FI} is
\begin{equation}\label{FIback}
\tilde{A}_\mu=0\,, \quad \tilde{\lambda}_{a\dot a}=0\,,\quad \tilde\Phi_{ab}=- 4\pi \zeta\, h_{ab}\,, \quad \tilde D_{\dot a\dot b}=\frac{4\pi \zeta}{r}\bar h_{\dot a \dot b}\,.
\end{equation}

\section{The current multiplet for standard gauge theories}\label{App:current}

We consider a $\mathcal{N}=4$ gauge theory for a single hypermultiplet ${\cal H} = \left( q_a, \tilde{q}^a, \psi_{\dot{a}},\tilde{\psi}_{\dot{a}} \right)$ with a weakly gauged $U(1)$ flavor symmetry. 

We construct the conserved current multiplet associated to the invariance of action \eqref{Shyper} under the global $U(1)$ symmetry 
\begin{align}
q_a&\to e^{\mathbbm{i}\alpha}q_a\,,  &\psi_{\dot a}&\to e^{\mathbbm{i}\alpha}\psi_{\dot a}\,,\\
\tilde q_a&\to e^{-\mathbbm{i}\alpha}\tilde q_a\,,  &\tilde\psi_{\dot a}&\to e^{-\mathbbm{i}\alpha}\tilde \psi_{\dot a}\,.
\end{align}
The corresponding Noether current is $j_\mu=\mathbbm{i}\tilde{q}^a\nabla_\mu q_a-\mathbbm{i}\nabla_\mu\tilde{q}^a\,q_a-\tilde{\psi}^{\dot b}\gamma_\mu\psi_{\dot b}$
and it sits in a current multiplet $\Sigma=\left(J_{ab},\,\chi_{a\dot a},\, \,j_\mu, \, K_{\dot a\dot b} \right)$. 

At linearized level, the multiplet is defined as the variation of the action w.r.t. the gauge multiplet $\Sigma\sim\frac{\delta S_{\textup{hyper}}}{\delta \mathcal{V}}$. Therefore, its components in terms of the hypermultiplet components are determined by comparing the linearized coupling
\begin{equation}
S_{\textup{lin}}=\int d^3x\sqrt{g}\left[A^\mu j_\mu+\mathbbm{i}D^{ab}J_{ab}+\Phi^{\dot a\dot b}K_{\dot a\dot b}+\lambda^{a\dot a}\chi_{a\dot a} \right]\,,
\end{equation} 
with the terms in $S_\textup{Hyper}$ linear in the vector field components. This allows to find the bottom component
\begin{equation}\label{bottom}
J_{ab}=-\tilde q_{(a}q_{b)}\,.
\end{equation}
In order to build the fully non-linear expressions of the higher components, we first observe that $S_{\textup{lin}}$ is SUSY invariant under the following variations of the current multiplet components  
\begin{align}
\delta_ \xi J_{ab}&=-\mathbbm{i}\xi\indices{_{(a}^{\dot a}}\chi_{b)\dot a}\,,\label{susy:curr1}\\
\delta_\xi\chi_{a\dot{a}}&=\xi\indices{_a^{\dot b}}K_{\dot a\dot b}+\gamma^\mu\xi\indices{^b_{\dot a}}\nabla_\mu J_{ab}+2\xi\indices{^{\prime\,b}_{\dot{a}}}J_{ab}+\frac{\mathbbm{i}}{2}\gamma^\mu\xi_{a\dot a}j_\mu \label{susy:curr2}\\
\delta_\xi K_{\dot a\dot b}&=-\mathbbm{i}\nabla_\mu\left(\xi\indices{^a_{(\dot a}}\gamma^\mu \chi_{|a|\dot b)}\right)-2\mathbbm{i} \xi\indices{^{\prime\,a}_{(\dot a}}\chi_{|a|\dot b)}\,,\label{susy:curr3}\\
\delta_\xi j_\mu&=\mathbbm{i}\epsilon_{\mu\nu\lambda}\xi^{a\dot a}\gamma^\nu\nabla^\lambda\chi_{a\dot a}-2\xi\indices{^{\prime \,a\dot{a}}}\gamma_\mu\chi_{a\dot a}\,.
\label{susy:curr4}
\end{align}
Now, applying $\delta_\xi$ to the bottom component \eqref{bottom} and using the explicit variations \eqref{hyper:var}, a comparison with \eqref{susy:curr1} allows to determine the $\chi_{a\dot a}$ fermion 
\begin{equation}
\chi_{a\dot a}=-\mathbbm{i}\left(\tilde{q}_a\psi_{\dot a}+\tilde{\psi}_{\dot a}q_a\right)\,.
\end{equation}
Comparing its variation obtained from \eqref{hyper:var} with variation \eqref{susy:curr2} we find the non-linear expressions for $j_\mu$ and $K_{\dot a\dot b}$
\begin{align}
K_{\dot a\dot b}&=\mathbbm{i}\tilde\psi_{(\dot a}\psi_{\dot b)}-\tilde q^b\Phi_{\dot a\dot b}q_b\\
j_\mu&=\mathbbm{i}\tilde{q}^aD_\mu q_a-\mathbbm{i}D_\mu\tilde{q}^a\,q_a-\tilde{\psi}^{\dot b}\gamma_\mu\psi_{\dot b}
\end{align}
Finally, it is easy to check that variations \eqref{hyper:var} imply that this multiplet correctly closes the algebra in (\ref{susy:curr1} - \ref{susy:curr4}).

\bibliography{biblio}

\end{document}